\documentclass[aps,twocolumn,superscriptaddress,altaffilletter,lengthcheck,tightenlines,showpacs,showkeys]{revtex4}

\newcommand{\ben}{\begin{eqnarray}}
\newcommand{\een}{\end{eqnarray}}
\newcommand{\be}{\begin{equation}}
\newcommand{\ee}{\end{equation}}
\newcommand{\ba}{\begin{eqnarray}}
\newcommand{\ea}{\end{eqnarray}}
\newcommand{\n}{\label}
\newcommand{\no}{\noindent}

\newcommand{\ga}{\gamma}
\newcommand{\ro}{\rho}

\newcommand{\bn}{\begin{equation}\label}
\newcommand{\ar}{${\cal A}^{^{R}}$}
\newcommand{\ai}{${\cal A}^{^{I}}$}
\newcommand{\am}{$a_{_{+}}$}
\newcommand{\ame}{$a_{_{-}}$}
\newcommand{\br}{${\cal B}^{^{R}}$}
\newcommand{\bi}{${\cal B}^{^{I}}$}
\newcommand{\cre}{${\cal C}^{^{R}}$}
\newcommand{\ci}{${\cal C}^{^{I}}$}
\newcommand{\plr}{${\cal P}^{^{R}}$}
\newcommand{\pli}{${\cal P}^{^{I}}$}

\usepackage[dvipdf]{epsfig}
\usepackage{color}
\usepackage{multirow}

\begin{document}

\title{Form invariance symmetry generates a large set of FRW cosmologies}

\author{Luis P. Chimento}\email{chimento@df.uba.ar}
\affiliation{Departamento de F\'{\i}sica, Facultad de Ciencias Exactas y Naturales,  Universidad de Buenos Aires, Ciudad Universitaria, Pabell\'on I, 1428 Buenos Aires, Argentina}
\author{Mart\'{\i}n G. Richarte}\email{martin@df.uba.ar}
\affiliation{Departamento de F\'{\i}sica, Facultad de Ciencias Exactas y Naturales,  Universidad de Buenos Aires, Ciudad Universitaria, Pabell\'on I, 1428 Buenos Aires, Argentina}
\author{Iv\'an E. S\'anchez G.}\email{isg@df.uba.ar}
\affiliation{Departamento de F\'{\i}sica, Facultad de Ciencias Exactas y Naturales,  Universidad de Buenos Aires, Ciudad Universitaria, Pabell\'on I, 1428 Buenos Aires, Argentina}

\date{\today}
\bibliographystyle{plain}

\begin{abstract}
We show that Einstein's field equations for spatially flat Friedmann-Robertson-Walker (FRW) space times have a form invariance symmetry (FIS) realized by the form invariance transformations (FIT) which are indeed generated by an invertible function of the source energy density. These transformations act on the Hubble expansion rate, the energy density, and pressure of the cosmic fluid; likewise such transformations  are endowed  with a Lie group structure. Each representation of this group is associated with a particular fluid and consequently a determined cosmology, so that, the FIS defines a set of equivalent cosmological models. We focus our seek in the FIT generated by a linear function because it provides a natural framework to express the duality and also produces a large sets of cosmologies, starting from a seed one, in several contexts as for instance in the cases of a perfect fluid source and a scalar field driven by a potential depending linearly on the scalar field kinetic energy density.


\end{abstract}

\date{\today}

\bibliographystyle{plain}

\maketitle

\section{Introduction}

Form invariance transformations involve internal or external variables in such a way that the transformations preserve the form of the dynamical equations, i.e., they have a form invariance symmetry  (FIS) \cite{symm}. Particularly useful are the T-duality \cite{Polchinsky(a)}, \cite{Polchinsky(b)} or ``scale-factor duality'' \cite{VET(a)}, \cite{VET(b)}. The former stems from quantum mechanics due to the possibility of interchanging the winding modes of a closed string with its momentum modes. T-duality relates a theory, which is compactified on a circle with radius $R$, with another compactified theory on a circle with radius $1/R$. The latter ones applies on the cosmological context and reflects the invariance property of the background dynamical equations. For a spatially flat FRW background the radius $R$ is replaced by the scale factor $a$, and the dual transformation $a\rightarrow a^{-1}$ connects a contracting cosmology with an expanding one.

A new kind of internal symmetry that preserves the form of the spatially flat Friedmann cosmology was found by one of the authors \cite{Luis}. There it was shown that the equations governing the evolution of  FRW cosmologies have a FIS group. The FIT which preserves the form of those equations relates quantities of the fluid, energy density and pressure, with geometrical quantities such as the scale factor and Hubble expansion rate. The FIS introduce an alternative concept of equivalence between different physical problems meaning that essentially a set of cosmological models are equivalent when their dynamical equations are form invariant under the action of some internal symmetry  group. This novel fact has motivated a large number of studies in FRW cosmologies for different cosmological setups.

A particular realization of the FIS is given by the transformation $\ro\to\ro$, thus it leads to $(i)-$ the identity $H\rightarrow H$ with $a\rightarrow a$ and $(ii)-$ the dual symmetry $H\rightarrow -H$ with $a\rightarrow 1/a$. This results also was relates with the additivity of the stress-energy tensor of the Einstein equations, namely, the different ways of summing two interacting fluid components induce the two above discrete symmetries \cite{Luis2}. It was noted in \cite{LCH} that the duality in the scalar field is equivalent to a Wick rotation. So, there is a duality between a scalar field cosmology and a phantom one \cite{phantomcosmo(a),phantomcosmo(b),phantomcosmo(c),phantomcosmo(d),phantomcosmo(e),phantomcosmo(f)}; this duality also was mentioned in \cite{indep}. This is interesting because one could be able to establish a link between phantom and standard scalar field cosmologies as it happens in the pre-big bang models \cite{GasVene}. Later on, it was found a method for obtaining some phantom k-essence cosmologies with the aid of the FIS \cite{phank}. Here, the phantom symmetry involves a change in the  potential which leads to an extended super-accelerated tachyon field \cite{phank}. This concept was extended in \cite{upc} where it was introduced ``the generalized duality'' $a\to a^n=1/a^{-n}$.  Duality was also investigated for the quintom cosmology \cite{CD1}, in the generalized quintom scenario characterized by $N$-scalar fields endowed with an internal metric \cite{quintomfis} and discussed providing a map between exact solutions corresponding to different values of the barotropic index \cite{Faraoni}. Further, in the latter it was investigated the role of the duality in the context of Brans-Dicke theory.

Actually the cooperative effects of increasing the energy density in the Friedmann equation make inflation more likely \cite{Luis}. This framework extends the assisted inflation mechanism relating it with the FIS of the FRW equations \cite{Luis}, \cite{otrosfis(a),otrosfis(b),otrosfis(c)}. Concerning this, two of the authors have introduced the enhanced inflation in the context of brane inflation for the Dirac-Born-infeld (DBI) theory \cite{LRM}. There, it was used the linear generating function $\rho \rightarrow \ga^2 \rho$ which increases the energy density of the DBI field because the ``Lorentz factor'' $\ga>1$. In particular, when $\ga$ is a natural number $n$, it plays the role of $n$ self-interacting DBI fields \cite{LRM}. Besides,  inflationary solutions in the induced modified gravity theory were obtained using FIS \cite{DGPFI}. Thus the FIS offers a nice interpretation of the duality and the enhanced inflation mechanism \cite{LRM} in terms of internal symmetries \cite{Luis}.  Another simple  procedure for generating new  inflationary  solutions  of  Einstein's equations  from old  ones was found  in \cite{Barrow} by  implementing a rescaling symmetry of the scalar field or constant scalings of the time variable. 

The FIS makes possible to find exact solutions in several contexts and generate new cosmologies from a known ``seed'' one  \cite{quintomfis}, \cite{CM1}. For instance, a unified phantom cosmology was obtained by applying a dual symmetry on the non-phantom one \cite{upc}, \cite{quintomfis},  \cite{LRM}. The general solution of  the $N$-quintom scalar field driven by a Liouville potential was found for a constant internal metric \cite{LRM}. Also it was noticed that k-essence cosmologies generated by a linear kinetic function and quintessence ones share the same scale factor when the $k$-field and the scalar field are driven by an inverse square potential and an exponential potential respectively \cite{Luis2}. On the other hand, the FIS has proved to be a  useful tool for  investigating different cosmological scenarios as fermion field \cite{fermionfis}, DBI field \cite{dbifis}, conformal scalar field \cite{Luis2}, \cite{zim}, phantom and inflationary set up with a decaying cosmological function for a 5D vaccum state \cite{1B}. Then we find interesting to investigate the consequences of the FIS from the physical and mathematical point of view.

Our main goal is to investigate the FIS of the Einstein equations for a FRW spacetime with several sources such as a perfect fluid and  a homogeneous scalar field, particularly when the latter is driven by a potential depending linearly on the scalar field kinetic energy density. We will discuss the main ideas behind the FIS and linear FIT which include the identity and dual transformation. After that, we will start with some seed cosmology  and use FIT to obtain a different new one. This will allows us to clarify some consequences derived from the Lie group structure of above transformations.

\section{FIS in flat FRW cosmology}

We will investigate an internal symmetry contained in the Einstein equations for a spatially flat FRW space-time
\be
\label{E1a}
3H^{2}=\rho\,
\ee
\be
\label{E1b}
\dot{\rho}+ 3H(\rho+p)=0\,
\ee
where $H=\dot a/a$ is the Hubble expansion rate and $a(t)$ is the scale factor. We assume that the universe is filled with a perfect fluid having  energy density $\rho$ and pressure $p$. The two independent Einstein equations have  three unknown quantities $(H, p, \rho)$, hence the system of equations (\ref{E1a})-(\ref{E1b}) has one degree of freedom. This allows to introduce FIT which involves those quantities,

\be
\n{tr}
\bar\rho=\bar\rho(\rho),
\ee
\be
\n{th}
\bar H=\left(\frac{\bar\rho}{\rho}\right)^{1/2}H,
\ee
\be
\n{tp+r}
\bar p+\bar\rho=\left(\frac{\rho}{\bar\rho}\right)^{1/2}\frac{d\bar\rho}{d\rho}\,\,(\rho+p).
\ee
They transform the corresponding barred Einstein equations (\ref{E1a})-(\ref{E1b}), with $\rho$, $p$ and $H$ replaced by $\bar{\rho}$, $\bar{p}$ and $\bar H$ into the same equations, here we have assumed that the function $\bar\rho(\rho)$ is invertible. Hence, the FIT (\ref{tr})-(\ref{tp+r}), generated by the function $\bar\rho(\rho)$, make the job of preserving the form of the system of equations (\ref{E1a})-(\ref{E1b}) and the FRW cosmology has a FIS. From the mathematical point of view, the FIS means that the FIT (\ref{tr})-(\ref{tp+r}) map solutions of the original system of equations, which indeed defines a cosmology through the variables $(H,p,\rho)$, into solutions of other system of equations,  defining  a different cosmology identified with the barred variables $(\bar{H}, \bar{p},\bar{\rho})$.

The FIT (\ref{tr})-(\ref{tp+r}) have a Lie group structure generated by the real invertible function $\bar{\rho}(\rho)$. The identity is induced by the transformation $\bar\ro=\ro$. The composition of  two consecutives FIT induced by $\bar\ro=\bar\ro(\ro)$ is associative,  so taking into account the  previous map one can readily write $\bar{\bar\ro}=\bar{\bar\ro}(\bar\ro)$ for the second composition,
\be
\n{comh}
\bar{\bar H}=\left(\frac{\bar{\bar\rho}}{\bar\rho}\right)^{1/2}\bar H=\left(\frac{\bar{\bar\rho}}{\rho}\right)^{1/2} H,
\ee
\be
\n{comp}
\bar{\bar p}+\bar{\bar\rho}=\left(\frac{\bar\rho}{\bar{\bar\rho}}\right)^{1/2}\frac{d\bar{\bar\rho}}{d\bar\rho}\,\,(\bar\rho+\bar p)=\left(\frac{\rho}{\bar{\bar\rho}}\right)^{1/2}\frac{d\bar{\bar\rho}}{d\rho}\,\,(\rho+p),
\ee
where it has used $\bar{\bar\ro}=\bar{\bar\ro}(\ro)$. Finally, the inverse  transformations of the FIT (\ref{th}) and (\ref{tp+r}), induced by $\ro=\ro(\bar\ro)$, are obtained from Eqs. (\ref{comh}) and (\ref{comp}) by taking $\bar{\bar\rho}=\ro$.

The FIS of the Einstein equations is evidenced through the FIT (\ref{tr})-(\ref{tp+r}), that in turn form a Lie group. The aforesaid symmetry shows an equivalence between the barred and the unbarred cosmologies.

\section{Linear FIT}

In this section we present FIT induced by the linear generating function $\bar\ro=n^2\ro$ being $n$ a constant. After this choice Eqs. (\ref{tr})-(\ref{tp+r}) become
\be
\n{trn}
\bar\ro=n^2\ro,
\ee
\be
\n{thn}
\bar H=nH,\quad \Rightarrow \quad \bar a=a^n,
\ee
\be
\n{tpn}
(\bar\ro+\bar p)=n(\ro+p).
\ee
Hence, the linear transformation (\ref{trn}) leads to a linear combinations of the variables $\ro, H, p$ and a power transformation of the scale factor, obtained after having integrated $\bar H=nH$. Finally, the Eq. (\ref{tpn}) gives the transformation rule for the pressure of the fluid
\bn{tp}
\bar p=-n^2\ro+n(\ro+p).
\ee

The linear FIT Eqs. (\ref{trn})-(\ref{tp}) have a Lie group structure. In fact,  for $n=1$ we have the identity $\bar\ro=\ro$, $\bar H=H$,  $\bar a=a$, and $\bar p=p$. The composition of the two transformations $\bar{\bar\rho}=\bar n^2\bar\ro$ and $\bar\ro=n^2\bar\ro$ gives $\bar{\bar\rho}=\bar n^2\bar\ro=\bar n^2 n^2\ro=\bar{\bar n}^2\ro$, with $\bar{\bar n}=\bar n n$. For the Hubble expansion rate and the pressure we get $\bar{\bar H}=\bar n\bar H=\bar n nH=\bar{\bar n}H$ and $\bar{\bar p}=-\bar n^2\bar\ro+\bar n(\bar\ro+\bar p)=-\bar n^2 n^2\ro+\bar n n(\ro+p)=-{\bar{\bar n}}^2\ro+{\bar{\bar n}}(\ro+p)$ while for the scale factor, we obtain $\bar{\bar{a}}= \bar{a}^{\bar{n}}=a^{\bar n n}=a^{\bar{\bar{n}}}$. Then the inverse transformation is found by making ${\bar{\bar n}=\bar n n=1}$, meaning that $\ro=\bar n^2\bar\ro$, $H=\bar n\bar H$, $ p=-\bar n^2\bar\ro+\bar n(\bar\ro+\bar p)$ and $a=\bar{a}^{\bar{n}}$.

In the case of considering two universes, each one of them filled with a perfect fluid for which we assume equations of state $\bar p=(\bar\ga-1)\bar\ro$ and $p=(\ga-1)\ro$ respectively,  the  barotropic index $\ga$ transforms as
\be
\n{tg}
\bar\ga=\frac{(\bar\ro+\bar p)}{\bar\ro}=\frac{\ro+p}{n\,\ro}=\frac{\ga}{n},
\ee
after using Eq. (\ref{trn}) along with Eq. (\ref{tpn}).

The existence of a Lie group structure opens the possibility of connecting the scale factor $a$ of a  seed cosmology with the scale factor $\bar a =a^n$ of a different cosmology. Below we will show several examples for getting a better understanding about this connection.

\subsection{Identity and duality}

A particularly simple transformation, is generated by the identity  $\bar\ro=\ro$ with $n^2=1$. From Eqs. (\ref{thn}) and (\ref{tpn}), it induces the transformations
\be
\n{ti}
\bar H=H, \qquad  \bar\ro+\bar p=\ro+p, \qquad  \bar a=a,
\ee
\be
\n{td}
\bar H=-H,  \qquad  \bar\ro+\bar p=-(\ro+p)  \qquad  \bar a=1/a.
\ee
The former, identical transformation, leads to the {\it identity} $\bar a=a$ and the latter, dual transformation, to the {\it duality} $\bar a=1/a$ with $\bar\ga=(\bar\ro+\bar p)/\bar\ro=-(\ro+p)/\ro=-\ga$ whereas its associated matter violates the weak energy condition $\ro+p<0$. There is a duality between expanding ($H>0$ and $\dot H<0$) and contracting ($H <0$ and $\dot H >0$) scenarios, and there is a duality between contracting ($H<0$ and $\dot H<0$) and super-accelerated expanding ($H >0$ and $\dot H >0$) cosmologies. In the latter case the energy density $\dot\ro=-3H(\ro+p)\ge 0$ is an increasing function of the time. In particular, if $\rho$ and $p$ diverge the dual transformation exchanges a final Big Crunch by a final Big Rip.

\subsection{The perfect fluid case}

To illustrate the main features of the Lie group,  represented by means of the FIS, we  begin by choosing a particular solution of the Eqs (\ref{E1a})-(\ref{E1b}) for a perfect fluid, ``seed solution'', and apply to it the FIT (\ref{trn})-(\ref{tp}) for obtaining a new set of solutions.  We adopt a constant equation of state $p=(\ga-1)\ro$ for the perfect fluid and select the seed solution corresponding to a matter dominated FRW universe, i.e., $\ga=1$ and $p=0$. So, the solutions of Eqs. (\ref{E1a})-(\ref{E1b}) are given by
\be
\n{scon}
\ro=\ro_0\left(\frac{a_0}{a}\right)^{3}, \qquad a=a_0\left[\sqrt{\frac{3\ro_0}{4}}\,\Delta t\right]^{2/3},
\ee
where $\ro_0$ is the energy density at $a=a_0$. Now, from Eq. (\ref{trn}) we obtain
\be
\n{ro0}
\bar\ro_0\left(\frac{\bar a_0}{\bar a}\right)^{3\bar\ga}=n^2\ro_0\left(\frac{a_0}{a}\right)^{3}.
\ee
Taking into account that $\bar a^{3\bar\ga}=a^{3n\bar\ga}=a^{3\ga}=a^{3}$ (see Eqs. (\ref{thn}) and (\ref{tg}) for $\ga=1$), the latter equation (\ref{ro0}) gives the transformation rule for the following combination of constants $\ro_0$ and $a_0^3$,
\be
\n{ro1}
\bar\ro_0\bar a_0^{3\bar\ga}=n^2\ro_0 a_0^{3},
\ee
which, combined with $\bar\ga=1/n$, becomes
\be
\n{tr0}
\bar\ga^2\bar\ro_0\bar a_0^{3\bar\ga}=\ro_0 a_0^3.
\ee
Then, using the transformations, $\bar\ro=\ro/\bar\ga^2$, $\bar a=a^{1/\bar\ga}$ and (\ref{tr0}) in the seed solution (\ref{scon}) we find the energy density and the scale factor of other cosmological model with a fluid characterized by the barotropic index $\bar\ga$,
\be
\n{a1f}
\bar\ro=\bar\ro_0\left(\frac{\bar a_0}{\bar a}\right)^{3\bar\ga}, \qquad \bar a=\bar a_0\left[\sqrt{\frac{3\bar\ga^2\bar\ro_0}{4}}\,\Delta t\right]^{2/3\bar\ga}.
\ee
Removing the bar, this scale factor becomes the general solution of the Friedmann equation for the source $\ro=\ro_0\left(a_0/{a}\right)^{3\ga}$.

\section{The scalar field}

Now, we turn our attention to the scalar field and will show how it transforms under the FIT  (\ref{trn})-(\ref{tpn}). We consider a self-interacting homogeneous scalar field $\phi=\phi(t)$ driven by a potential $V (\phi)$, so the Einstein-Klein-Gordon (EKG) equations are
\be
\label{A}
3H^2=\frac{1}{2}\dot{\phi}^2+V,
\ee
\be
\n{kg}
\ddot\phi+3H\dot\phi+\frac{dV}{d\phi}=0.
\ee
Using the perfect fluid interpretation of the scalar field, through the standard identifications
\be
\rho_\phi=\frac{1}{2}\dot{\phi}^2+V, \qquad \quad p_\phi=\frac{1}{2}\dot{\phi}^2-V,  \label{roE}
\ee
\be
\n{gs}
\ga_\phi=\frac{\dot\phi^2}{\ro},
\ee
we find that the  scalar field kinetic energy density, the potential and the barotropic index transform linearly under the FIT (\ref{trn})-(\ref{tpn}),
\be
\label{tf}
\dot{\bar{\phi}}^{2} =n\dot{\phi}^{2},
\ee
\be
\bar{V}=\frac{n(n-1)}{2}\,\dot{\phi}^2+n^2 V,  \label{tv}
\ee
\be
\n{tgs}
\bar\ga_\phi=\frac{\ga_\phi}{n},
\ee
and the scalar field transforms as $\bar\phi=\sqrt{n}\,\phi$.

{Here, we  will discuss some interesting points regarding the transformation rules  for the scalar field, potential, and barotropic index  (\ref{tf})-(\ref{tgs}). To provide a better understanding of the mix of terms contained into Eq. (\ref{tv}), like the kinetic energy density $\dot\phi^2$ and the potential $V(\phi)$, we choose the class of potentials which can be written as a function of the kinetic energy density, $V(\phi)=F(\dot{\phi}^{2})$. It means that $dV/d\phi=2\ddot\phi\,dF/d\dot\phi^2$ so that the KG equation (\ref{kg}) becomes
\be
\n{kg1}
\left(1+2\frac{dF}{d\dot\phi^2}\right)\ddot\phi+\dot\phi {\sqrt{3\left(\frac{\dot{\phi}^2}{2}+F\right)}}=0,
\ee
while its first integral
\be
\n{kg11}
\phi=-\int{\frac{1+2\frac{dF}{d\dot\phi^2}}{{\sqrt{3\left(\frac{\dot{\phi}^2}{2}+F\right)}}}\,\,d\dot\phi},
\ee
expresses the dependence of the scalar field on $\dot{\phi}^{2}$ and consequently the dependence of the potential on $\phi$, through   $V(\phi)=F(\dot{\phi}^{2})$, after inverting  the integral (\ref{kg11}). The same goes for the barred quantities and the class of potential having the form $\bar V(\bar\phi)=\bar F(\dot{\bar\phi}^2)$. The relation between the barred and unbarred scalar fields is given by Eq. (\ref{tf}). Finally, it should be stressed that  form invariance symmetry is an internal symmetry of Einstein field equations but it is not  a symmetry of the gravitational Lagrangian minimally coupled to matter fields, where  the well-known Noether's theorem applies \cite{NT1},  \cite{NT2}.}

\subsection{Potential depending linearly on $\dot\phi^2$ }

Let us consider a scalar field driven by a potential depending linearly on the scalar field kinetic energy density,
\be
V(\phi)=\frac{2-\ga_0}{2\ga_0}\,{\dot\phi}^{2}+V_0,  \label{VL}
\ee
where the parameters $\ga_0$ and $V_0$ are two free parameters and $\dot{\phi}=\dot{\phi}(\phi)$.
{This parametrization of the potential is really useful   because it involves a linear function of kinetic energy density, hence it preserves this ``linearity" and its form when it is transformed by a FIT (\ref{tf})-(\ref{tgs}) to the barred cosmology,
\be
\bar V(\bar\phi)=\frac{2-\bar\ga_0}{2\bar\ga_0}\,{\dot{\bar\phi}}^{2}+\bar V_0.  \label{VL'}
\ee
In addition, for the potential (\ref{VL}) the function $F(\dot{\phi}^{2})=(2-\ga_{0})\dot{\phi}^{2}/2\ga_{0}+V_{0}$ is also linear, so that the Eq. (\ref{kg11}) becomes
\bn{kgk}
\phi=-\frac{2}{\ga_0}\int\frac{d\dot\phi}{\sqrt{3\left(\frac{\dot{\phi}^2}{\ga_0}+V_0\right)}},
\ee
which is easily integrable and invertible. After integrate the latter equation we  obtain  $\phi=\phi(\dot\phi^2)$ and consequently $\dot\phi^2=\dot\phi^2(\phi)$ after invert it. The property that Eq. (\ref{kgk}) be integrable and invertible is very important because it assures the reconstruction process of getting the potential as a function of the scalar field through  $V(\phi)=F(\dot{\phi}^{2}(\phi))$. In the next subsections we will use $\dot\phi^2(\phi)$ to express the potential (\ref{VL}) as a function of the scalar field. }

For the potential (\ref{VL}), $dV/d\phi=\ga^{-1}_0(2-\ga_0)\ddot\phi$ and the EKG equations (\ref{A})-(\ref{kg}) reduce to
\bn{00s}
3H^2=\frac{c^2}{\ga_0 a^{3\ga_0}}+V_0,
\ee
\be
\n{pi}
\dot\phi^2=\frac{c^2}{a^{3\ga_0}},
\ee
where the latter equation is the first integral of the Klein-Gordon (KG) equation (\ref{kg}), $c^2$ is an integration constant and for $c^2<0$ the  scalar field is imaginary.
{The latter case corresponds to have a phantom field with a negative kinetic energy density given by $-\dot{\phi}^{2}/2$, this scenario was  widely explored in the literature \cite{LCH}, \cite{phantomcosmo(a),phantomcosmo(b),phantomcosmo(c),phantomcosmo(d),phantomcosmo(e),phantomcosmo(f)}, \cite{upc}, \cite{quintomfis} . It was obtained from the form invariance symmetry by taking $n=-1$ in the transformation rules (\ref{tf})-(\ref{tv}) for the kinetic energy density and potential. These rules imply that $\bar{V}\rightarrow V+\dot{\phi}^{2}$, $\dot{\bar{\phi}}^{2}\rightarrow -\dot{\phi}^{2}$ and $\bar{\rho}_{\phi}=\rho_{\phi}$, while the Hubble expansion rate changes as $\bar{H}=-H$ with $\bar a\rightarrow 1/a$ and $\bar\ga_\phi\rightarrow -\ga_\phi$. Finally the EKG equations remains invariant, meaning that the RHS of Friedmann equation of the barred cosmology $3\bar H^2=\bar\rho_\phi=\ro_\phi>0$ includes a negative kinetic energy density. In spite of that, the phantom scalar field does not introduce any difficulty in the model because we always obtain a real scalar factor.}

In other words, the linear potential  (\ref{VL}) describes the well-known flat FRW universe filled with a perfect fluid, whose equation of state is given by $p=(\ga_0-1)\ro$, and having a cosmological constant $V_0$.

From Eqs. (\ref{pi}) and  (\ref{VL}), we have $\dot{\phi}=\dot{\phi}(a)$ and $\phi=\phi(a)$.

The exact form of the scale factor and the scalar field are easily obtained by integrating the EKG equations (\ref{00s})-(\ref{pi}). For this reason it appears to be a good example to illustrate how from a seed solution, characterized by particular values of the parameters $V_0$, $\ga_0$ and $c^2$, the FIS helps us to find the scale factor and the scalar field driven by the linear potential (\ref{VL}) for any other value of those parameters. In this direction we will obtain the  transformation rules of those parameters under the FIT (\ref{trn})-(\ref{tpn}). Then, starting from
\be
\n{trr}
\bar\ro=\frac{{\dot{\bar\phi}}^2}{\bar\ga_0}+\bar V_0=n^2\ro=n^2\left(\frac{{\dot\phi}^{2}}{\ga_0}+V_0\right),
\ee
we obtain a relation that must be satisfied identically so, one gets ${\dot{\bar\phi}^2}/\bar\ga_0=n^2{\dot\phi}^{2}/\ga_0$ and $\bar V_0=n^2 V_0$. Then with the aid of the Eq. (\ref{tf}), the transformation rules for $\ga_0$ and $V_0$ are
\be
\n{tgv}
\bar\ga_0=\frac{\ga_0}{n},    \qquad  \quad \bar V_0=\frac{\ga_0^2}{\bar\ga_0^2}\, V_0.
\ee
Using the transformations (\ref{thn}), (\ref{tf}), and (\ref{tgv}) into Eq. (\ref{pi}),
\be
\n{tpi}
\dot{\bar{\phi}}^{2}=\frac{\bar c^2}{\bar a^{3\bar\ga_0}}=\frac{\bar c^2}{a^{3\ga_0}}=n\dot\phi^2=\frac{\ga_0}{\bar \ga_0}\frac{c^2}{a^{3\ga_0}},
\ee
we find the transformation rule for the parameter $c^2$,
\be
\n{tc2}
\bar c^2=\frac{\ga_0}{\bar\ga_0}c^2.
\ee
Combining the latter equation (\ref{tc2}) with the second formula of Eq. (\ref{tgv}) one gets a form invariant relation between the parameters $\ga_0$, $V_0$ and $c^2$
\be
\n{gvc}
\frac{\bar\ga_0 \bar V_0}{\bar c^2}=\frac{\ga_0 V_0}{c^2},
\ee
that will be very useful later on.



\subsection{The seed solution $V=0$}



{Setting  $\ga_0=2$ and $V_0=0$ in the potential (\ref{VL}), it vanishes and the corresponding EKG equations (\ref{00s})-(\ref{pi}) become
\bn{00s0}
3H^2=\frac{c^2}{2a^{6}},  \qquad  \dot\phi^2=\frac{c^2}{a^{6}}.
\ee
The solution of the first equation (\ref{00s0}) is given by
\bn{solv0}
a_{_{\mp}}=\left[\mp\sqrt{\frac{3c^2}{2}}\,\,t\right]^{1/3},
\ee
where we have taken advantage of the time translational invariance of the scale factor, which means that $a(t)=a(t-t_0)$, and set the initial singularity at $t=0$. The branches $a_{_{-}}$ and  $a_{_{+}}$ are defined for $t<0$ and $0<t$ while the final or initial singularity of the scale factor $a_{_{\mp}}$ is at $t=0$. Now, coming back to the first integral of the EKG equation, see the second equation (\ref{00s0}), and combining with the scale factor (\ref{solv0}), we have $\dot\phi^2=2/3t^2$, which after integrate it reads
\bn{solv00}
\phi=\sqrt{\frac{2}{3}}\,\ln{|t|},
\ee
Using  (\ref{solv00}) along with $\dot\phi^2=2/3t^2$, we obtain
\be
\n{.ff}
\dot\phi^2=\frac{2}{3}\, e^{-\sqrt{6}\,\phi}.
\ee}
Here we also used the field-translational invariance $\phi \rightarrow \phi -\phi_{0}$ to set $\phi_0=0$. So that $\phi$ vanishes at $|t|=1$. Resuming,
we have obtained a seed solution (\ref{solv0})-(\ref{.ff}) of the EKG equations (\ref{00s0}) representing to V = 0 and included in the set \plr, see Table (\ref{I}).

Applying the FIT (\ref{trn})-(\ref{tpn}) to the seed solution (\ref{solv0})-(\ref{.ff}) and using the Eqs. (\ref{tf}), (\ref{tv}), (\ref{tgv}) and (\ref{tc2}) we deduce that $\bar\ga_0=2/n$. Then, the scale factor and the scalar field of the barred cosmology are given by
\be
\n{sa0}
\bar a_{_{\mp}}=a_{_{\mp}}^n=\left[\mp\sqrt{\frac{3\bar\ga_0\bar c^2}{4}}\,t\right]^{2/3\bar\ga_0},
\ee
\be
\n{sf0}
\bar\phi=\sqrt{n}\,\phi=\frac{2}{\sqrt{3\bar\ga_0}}\,\ln{|t|}.
\ee
Combining Eqs. (\ref{tf}) and (\ref{.ff}) we can write $\dot\phi^2=\dot\phi^2(\bar\phi)$, so that $\bar\phi$ is driven by the potential
\be
\n{sv0}
\bar V=\frac{n(n-1)}{2}\,\dot{\phi}^2=\frac{2(2-\bar\ga_0)}{3\bar\ga_0^2}\,
\,e^{-\sqrt{3\bar\ga_0}\,\bar\phi}.
\ee
Now, removing the bar in Eqs. (\ref{sa0})-(\ref{sv0}) we obtain the well known power law solution (\ref{sa0}), the scalar field (\ref{sf0}) and the exponential potential (\ref{sv0}).{ There is a map between two cosmologies, driven by two different potentials each of which parametrized by $\ga_0$ and $\bar\ga_0=\ga/n$, when the FIT (\ref{trn})-(\ref{tpn}) is applied. In this sense, we have started with a vanishing potential, obtained for $\ga_{0}=2$ and $V_{0}=0$ in Eq. (\ref{VL}), and the transformed  scalar cosmology is driven by the exponential potential (\ref{sv0}) or between two different exponential potentials when the seed solution is found  on the case with $\ga_{0}\ne 2$ and $V_{0}=0$. }

The solution (\ref{sa0}) generalizes the assisted inflation because inflation occurs for $2<3\bar\ga_0$ or  $3<n$, where $n$ is the number of scalar fields, while in our framework $n$ is a real number and  the inflation is achieved by the cooperative effects of adding energy density into the Friedmann equation (\ref{00s}) instead of adding scalar fields, {see appendix A for more details. }
We denote with ${\cal P}$ the set of solutions (\ref{sa0})-(\ref{sf0}) for the potential (\ref{sv0}) [see Table (\ref{I})]. It includes the subset \plr, with $\ga_0>0$ and real scalar fields $c^2>0$, and the subset \pli, with $\ga_0<0$ and imaginary scalar fields $c^2<0$.
For $0<\ga_0< 2$, the power-law solution \ame represents a contracting universe ending in a Big Crunch at $t=0$, \am describe an expanding universe with a Big Bang at $t=0$ and the potential is positive. For $2<\ga_0$, the potential is negative but the scalar field and the scale factor retain the same features  mentioned above. However, for $\ga_0<0$ and $c^2<0$ things are quite different because, although the potential keeps positive, the scalar field becomes imaginary, the power law solution \ame describes an expanding universe ending in a  Big Rip at $t=0$ and \am represents a singular contracting universe. In all cases, the potential and the scalar field diverge at $t=0$ while the scale factor has an initial or final singularity at $t=0$.

\begin{center}
\begin{table}
\begin{tabular}{|l|l|l|}
\hline
\multicolumn{3}{|c|}{Classification}\\
\hline
\multirow{2}{*}{\,\,\,$V_{0}=0,\,\,\,\gamma_{0}c^{2}>0$\,\,\,}&\,\,\,\,\,\,${\cal P}$\,\,\,\,\,\, &\,\,\,\,\,\, \plr: $\gamma_{0}>0$,~~${\dot{\phi}}^2>0$\,\,\,\,\,\,\\
�&&\,\,\,\,\,\, \pli: $\gamma_{0}<0\,$,~~$\,{\dot{\phi}}^2<0$\,\,\,\,\,\, \\
\hline
\hline
\multirow{2}{*}{\,\,\,$V_{0}>0,\,\,\,\gamma_{0}c^{2}>0$\,\,\,}&\,\,\,\,\,\,${\cal A}$\,\,\,\,\,\, &\,\,\,\,\,\,\ar: $\gamma_{0}>0$,~~${\dot{\phi}}^2>0$\,\,\,\,\,\, \\
�&&\,\,\,\,\,\, \ai: $\gamma_{0}<0\,$,~~$\,{\dot{\phi}}^2<0$\,\,\,\,\,\, \\
\hline
\multirow{2}{*}{\,\,\,$V_{0}>0,\,\,\,\gamma_{0}c^{2}<0$\,\,\,}&\,\,\,\,\,\,${\cal B}$\,\,\,\,\,\, &\,\,\,\,\,\,\bi: $\gamma_{0}>0$,~~${\dot{\phi}}^2<0$\,\,\,\,\,\, \\
&&\,\,\,\,\,\, \br: $\gamma_{0}<0$,~~${\dot{\phi}}^2>0$\,\,\,\,\,\, \\
\hline
\multirow{2}{*}{\,\,\,$V_{0}<0,\,\,\,\gamma_{0}c^{2}>0$\,\,\,}&\,\,\,\,\,\,${\cal C}$\,\,\,\,\,\, &\,\,\,\,\,\,\cre: $\gamma_{0}>0$,~~${\dot{\phi}}^2>0$\,\,\,\,\,\, \\
&&\,\,\,\,\,\,\ci: $\gamma_{0}<0$,~~${\dot{\phi}}^2<0$\,\,\,\,\,\, \\
\hline
\end{tabular}
\caption{\label{I} Solutions of the EKG equations (\ref{00s})-(\ref{pi}) are  splitting  into four sets according to the values of  $V_0$, $\ga_0$, and $c^2$. $R$ and $I$ indicate real and imaginary scalar field respectively.}
\end{table}
\end{center}

\subsection{The seed solution $V=V_0$}

Now we will answer the following question: how could we proceed, to find the general solution of the EKG equations (\ref{00s})-(\ref{pi}) when the scalar field is driven by the potential (\ref{VL}) and we know only a particular solution of those equations? The answer will come from applying FIT on that particular solution. This  seed solution will be obtained by solving the Eqs. (\ref{VL})-(\ref{pi}) for $\ga_0=2$,
\bn{00a}
V=V_0, \qquad 3H^2=\frac{c^2}{2\,a^{6}}+V_0, \qquad \dot\phi^2=\frac{c^2}{a^{6}}.
\ee
It can be split into three different kinds of seed solutions as we choose $V_0>0$ and $c^2>0$, $V_0>0$ and $c^2<0$ or $V_0<0$ and $c^2>0$ which are accommodated into three subsets \ar, \bi and \cre, see Table (\ref{I}). The remaining solutions contained in the subsets \ai, \br and \ci, will be generated by applying the transformation rules (\ref{tgv}) and (\ref{tc2}) on those seed solutions. They map solutions of each one of the sets ${\cal A}=\,$\ar$\cup$\,\ai, ${\cal B}=\,$\br$\cup$\,\bi, ${\cal C}=\,$\cre$\cup$\,\ci into itself and each set will be associated with three different potentials as we will see below. Solving the Eqs. (\ref{00a}), we find the three seed solutions which are listed below,
\begin{itemize}
	\item (1) Seed solution included in \ar
\end{itemize}
\bn{aas}
a_{_{\mp}}=\left[\mp\sqrt{\frac{c^2}{2V_0}}\sinh{\sqrt{3V_0}}\,\,t\right]^{1/3},
\ee
\bn{pa}
\phi=\sqrt{\frac{2}{3}}\,\ln{\left|\tanh{\frac{\sqrt{3V_0}}{2}\,\,t}\right|},
\ee
\bn{p.2a}
\dot{\phi}^2=2V_{0}\sinh^2{\sqrt{\frac{3}{2}}\,\phi},
\ee
where the branches $a_{_{\mp}}$ are defined for $t^-<0$ and $0< t^+$ respectively, the final or initial singularity of $a$ was set at $t=0$ and $\phi$ was chosen to vanish at $|t|=\infty$.
\begin{itemize}
	\item (2) Seed solution included in \bi
\end{itemize}
\bn{ab}
a=\left[\sqrt{\frac{-\,c^2}{2V_0}}\cosh{\sqrt{3V_0}}\,\,t\right]^{1/3},
\ee
\bn{pb}
\phi=i2\sqrt{\frac{2}{3}}\,\arctan{\exp{\sqrt{3V_0}}\,\,t},
\ee
\bn{p.2b}
\dot{\phi}^2=-2V_{0}\sin^2{\sqrt{\frac{3}{2}}\,i\phi}.
\ee
The scale factor (\ref{ab}) is free of singularities, bounces at $t=0$ and the scalar field (\ref{p.2b}) is imaginary.
\begin{itemize}
	\item (3) Seed solution included in \cre
\end{itemize}
\bn{ac}
a=\left[\sqrt{\frac{-\,c^2}{2V_0}}\sin{\sqrt{-3V_0}}\,\,t\right]^{1/3},
\ee
\bn{pc}
\phi=\sqrt{\frac{2}{3}}\,\ln{\tan{\frac{\sqrt{-3V_0}}{2}}\,\,t},
\ee
\bn{p.2c}
\dot{\phi}^2=-2V_{0}\cosh^2{\sqrt{\frac{3}{2}}\,\phi}.
\ee
In this case the seed solution represents a universe with a finite time span where $0\le t\le \pi/\sqrt{-3V_0}$. Eqs. (\ref{p.2a}), (\ref{p.2b}) and (\ref{p.2c}) were obtained by integrating Eq. (\ref{kgk}).

\subsubsection{Generating the set of solutions ${\cal A}$}

The transformation rules for the parameters $\ga_0, V_0, c^2$ (\ref{tgv}) and (\ref{tc2})-(\ref{gvc}) evaluated on $\ga_0=2$ turn into
\bn{ns}
n=\frac{2}{\bar\ga_0}, \qquad V_0=\frac{\bar\ga_0^2\bar{V_0}}{4},
\ee
\bn{tcs}
c^2=\frac{\bar{\ga_0}\bar{c}^2}{2}, \qquad \frac{c^2}{2V_0}=\frac{\bar{c}^2}{\bar\ga_0\bar{V_0}}.
\ee
Inserting these unbarred parameters into the seed solution (\ref{aas})-(\ref{p.2a}) and using the  transformation rules for the scale factor (\ref{thn}), the scalar field (\ref{tf}) and the potential (\ref{tv}), we find the corresponding barred quantities
\be
\bar a_{_{\mp}}=a_{_{\mp}}^n=\left[\mp{\sqrt{\frac{\bar{c}^2}{\bar{\ga_{0}}\bar{V_{0}}}}}\,\,\sinh{\frac{\sqrt{3\bar V_{0}{\bar\ga_0}^2}}{2}}\,\,t\right]^{2/3\bar\ga_0}, \label{aa}
\ee
\be
\n{bsf2}
\bar\phi=\sqrt{n}\,\phi=\frac{2}{\sqrt{3\bar\ga_0}}\,\ln{\left|\tanh{\frac{\sqrt{3\bar V_{0}{\bar\ga_0}^2}}{4}\,\,t}\right|},
\ee
\be
{\dot{\bar{\phi}}}^2=n\dot\phi^2=\bar{V_0}\bar{\ga_0}\sinh^2{\frac{\sqrt{3\bar{\ga_0}}}{2}\,\,\bar\phi},  \label{bV>01}
\ee
\be
\bar V=\bar{V_0}\left[\cosh^2{\frac{\sqrt{3\bar\ga_0}}{2}\,\,\bar{\phi}}-\frac{\bar{\ga_0}}{2}
\sinh^2{\frac{\sqrt{3\bar\ga_0}}{2}\,\,\bar{\phi}}\right].  \label{bVL}
\ee
Removing the bar in Eqs. (\ref{aa})-(\ref{bVL}), we obtain the general solution of the system of equations (\ref{00s})-(\ref{pi}) or the ``solutions" of the EKG equations (\ref{A})-(\ref{kg}) driven by the potential (\ref{bVL}), for which the expansion rate can be written as a function of the scalar field $H=H(\phi)$. Note that $\sqrt{\bar\ga_0}\bar\phi=\sqrt{\ga_0}\phi$ is an invariant quantity.

\begin{figure}[hbt!]
\begin{minipage}{1\linewidth}
\resizebox{2.4in}{!}{\includegraphics{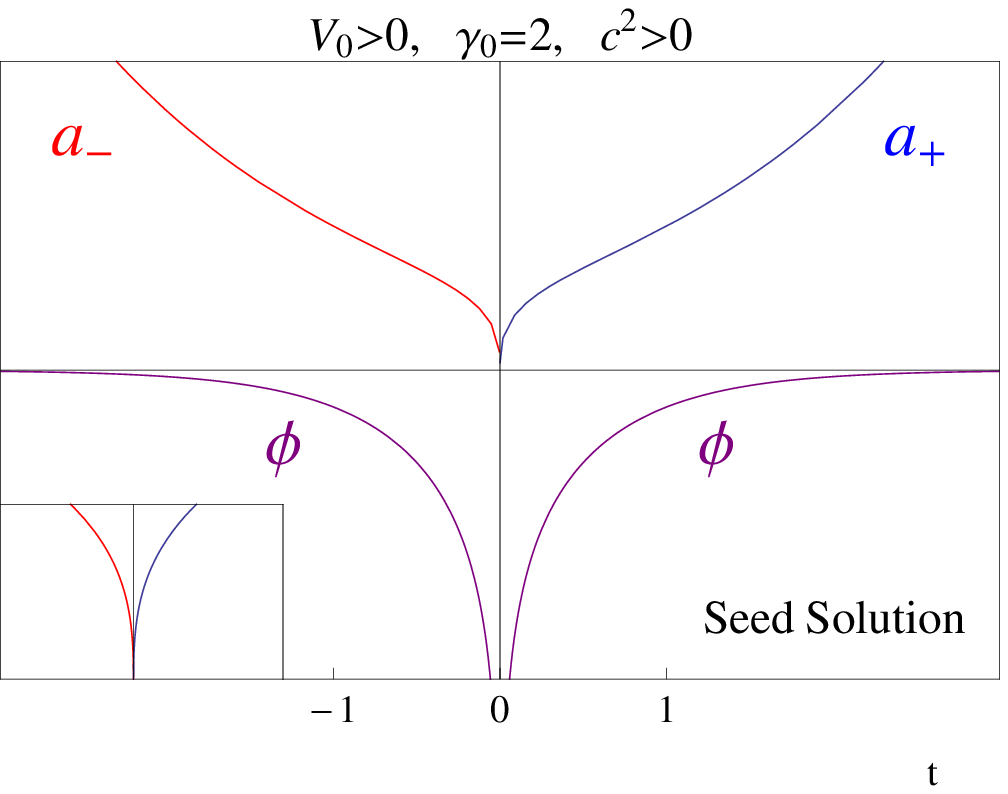}}\hskip0.05cm
\resizebox{2.4in}{!}{\includegraphics{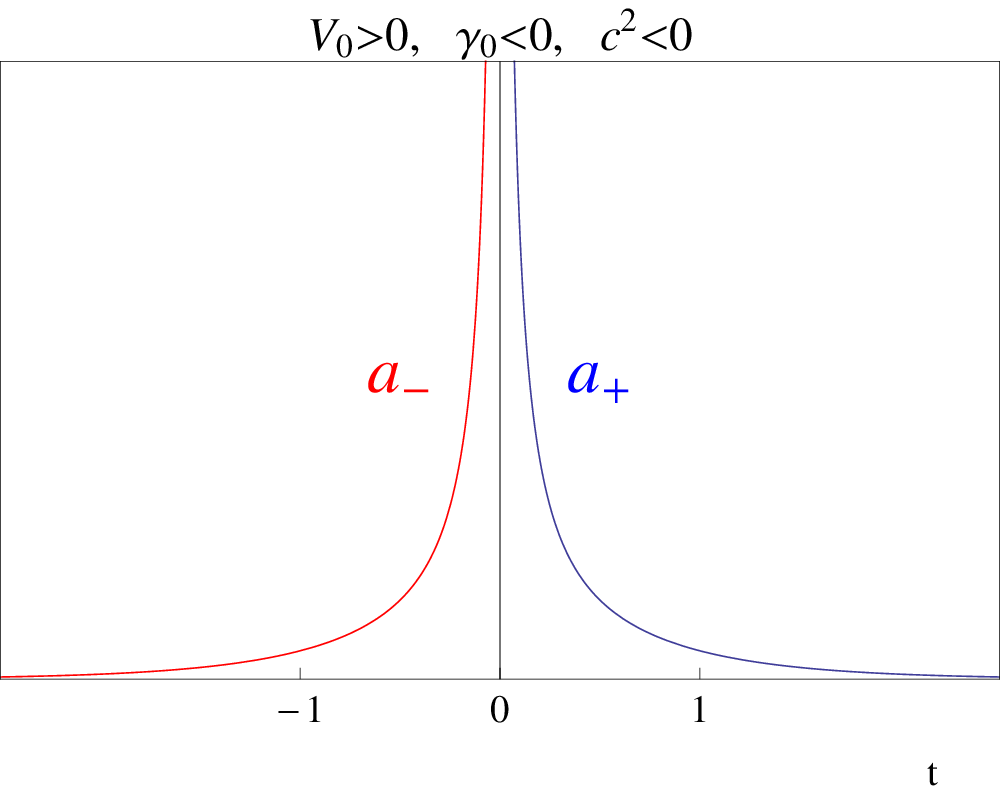}}\hskip0.05cm
\resizebox{2.4in}{!}{\includegraphics{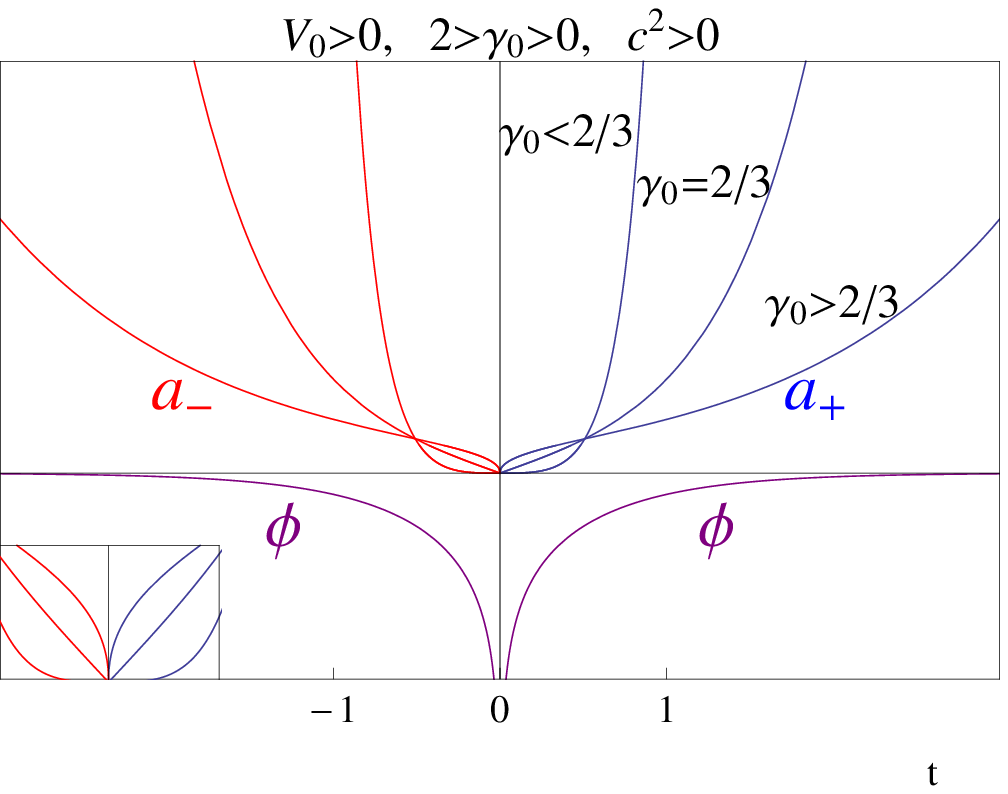}}\hskip0.05cm
\resizebox{2.4in}{!}{\includegraphics{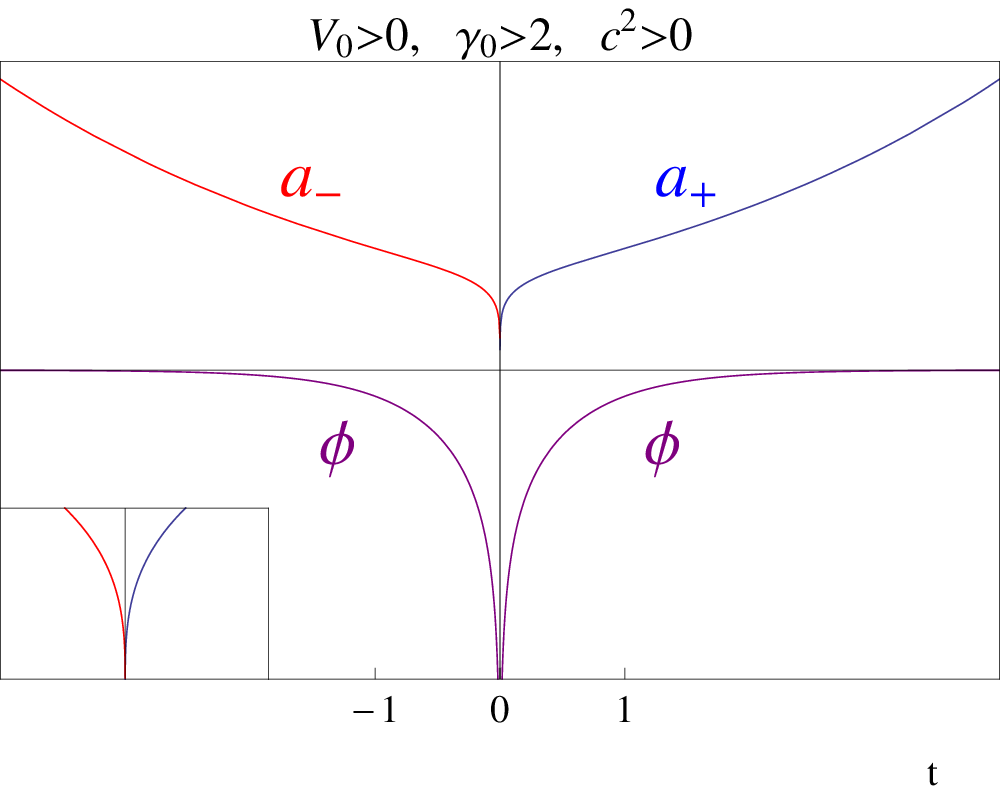}}\hskip0.05cm
\caption{\scriptsize{We plot $a(t)$ and $\phi(t)$ of the seed cosmology included in ${\cal A}^{R}$ and the remaining cosmologies obtained by applying the FIT.  We also plot $a(t)$ near the initial singularity.}}
\label{F1}
\end{minipage}
\end{figure}

The FIT connect two different cosmological scenarios, one  with $V_0>0$, $\ga_0>0$, $c^2>0$ and a real scalar field belonging to \ar, and another with $V_0>0$, $\ga_0<0$, $c^2<0$ and an imaginary scalar field belonging to \ai,
generating the whole set ${\cal A}=\,$\ar$\cup$\,\ai. Hence, the cosmological models associated with  solutions belonging to the set ${\cal A}$ are all related between them under FIT, for instance, a contracting universe ending in a Big Crunch at $t=0$, described by the branch $a_{_{-}}$, and a super-accelerated one ending in a Big Rip at $t=0$ or a singular expanding universe with a Big Bang at $t=0$, described by the branch $a_{_{+}}$ (\ref{aa}), and singular contracting one with an initial singularity at $t=0$ [see Fig. (\ref{F1})].

\begin{figure}[hbt!]
\begin{minipage}{1\linewidth}
\resizebox{2.4in}{!}{\includegraphics{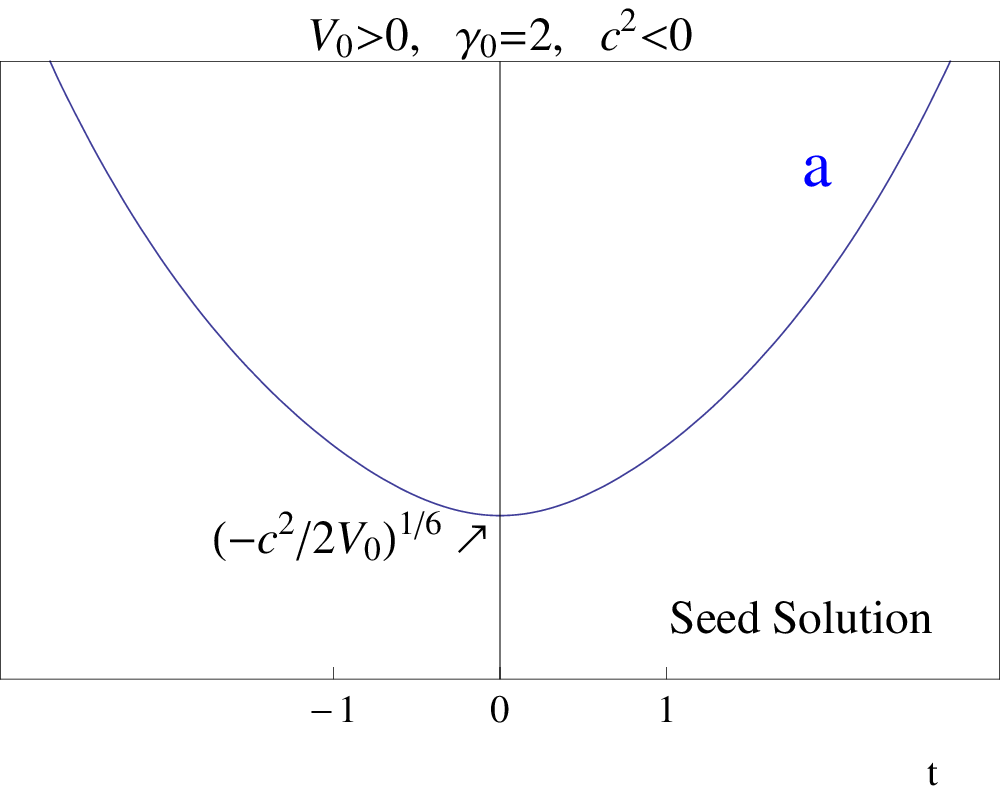}}\hskip0.05cm
\resizebox{2.4in}{!}{\includegraphics{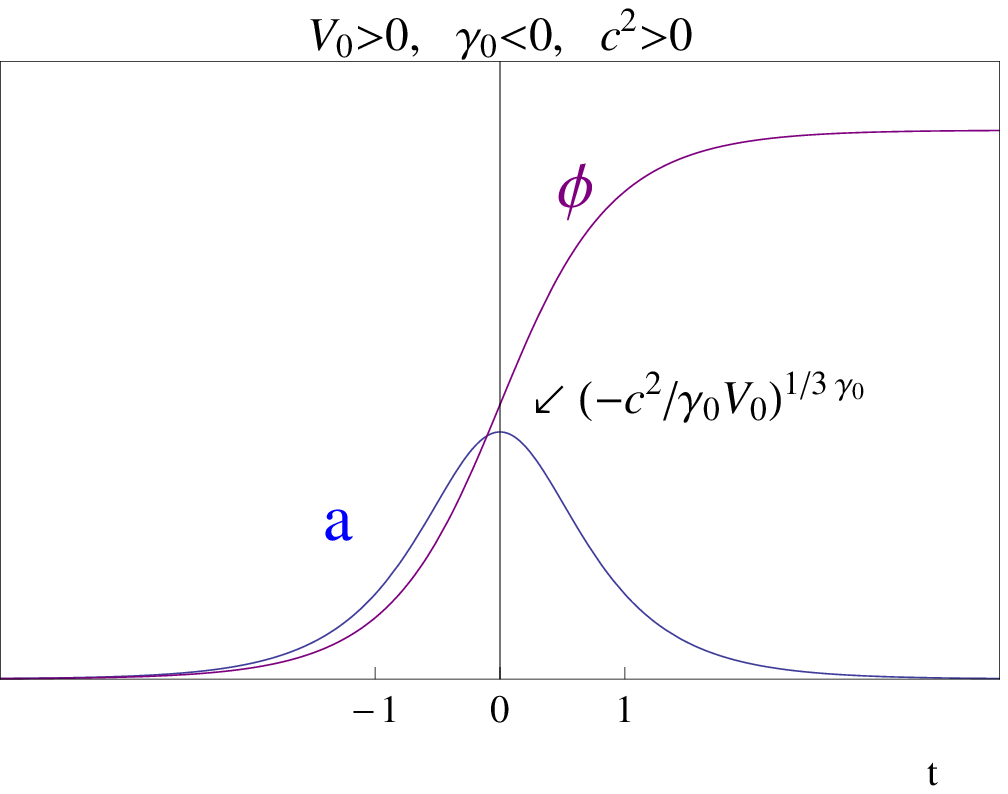}}\hskip0.05cm
\resizebox{2.4in}{!}{\includegraphics{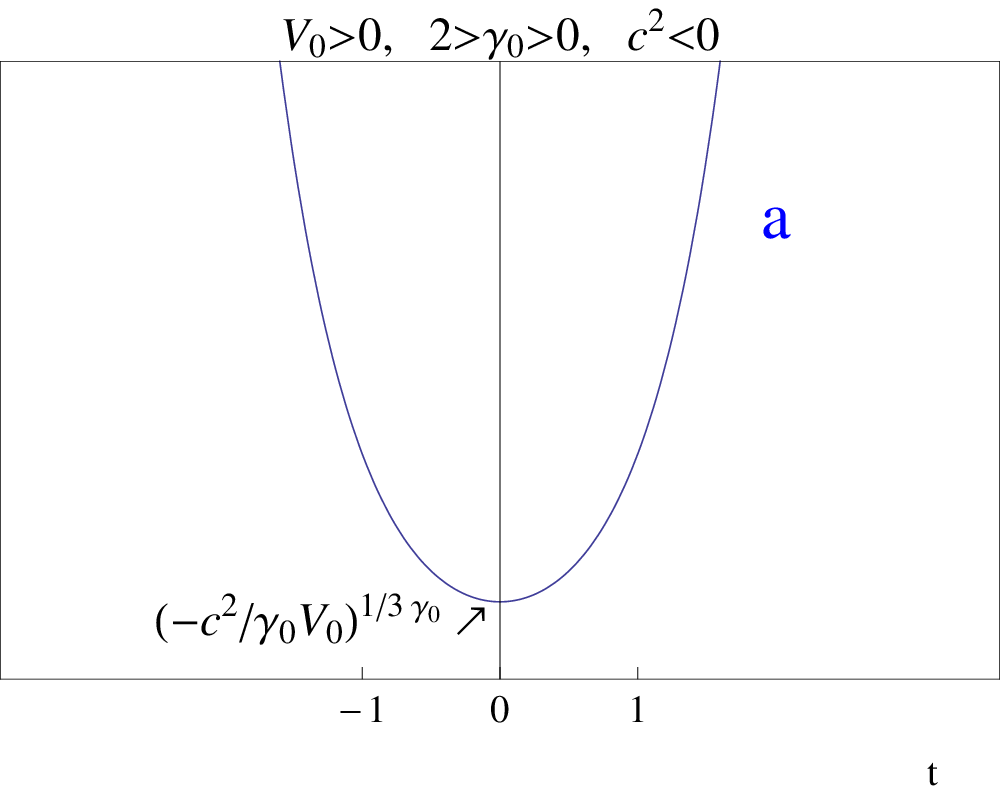}}\hskip0.05cm
\resizebox{2.4in}{!}{\includegraphics{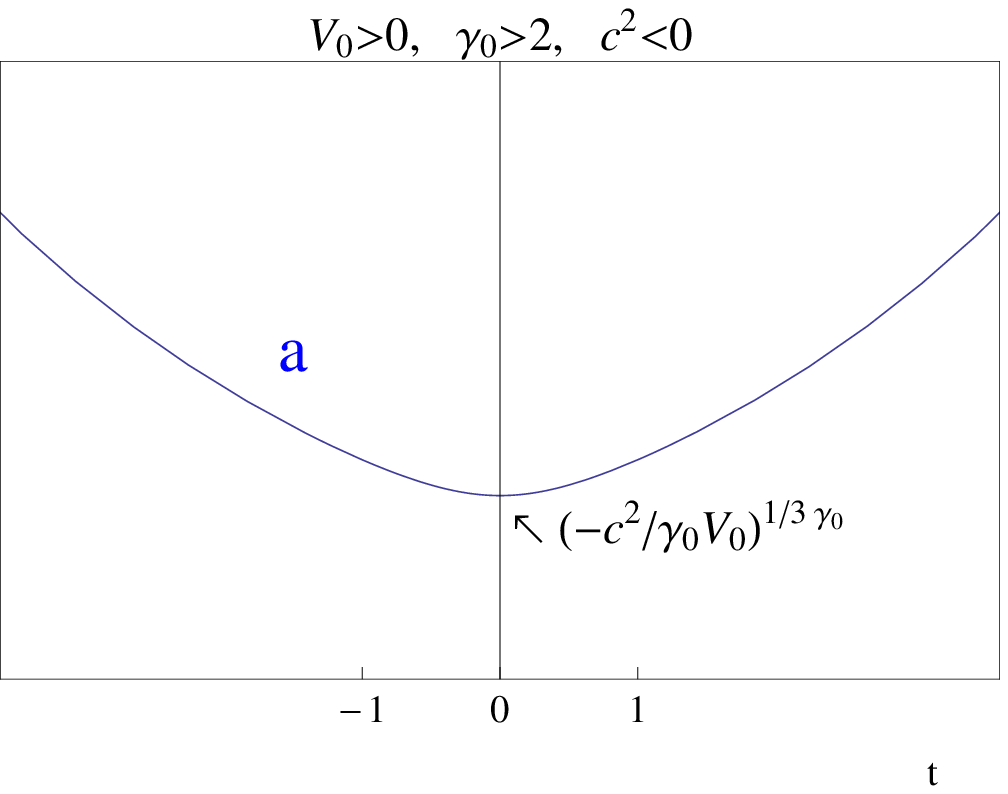}}\hskip0.05cm
\caption{\scriptsize{We plot $a(t)$ and $\phi(t)$ of the seed cosmology included in ${\cal B}^{I}$ and the remaining cosmologies obtained by applying the FIT.}}
\label{F2}
\end{minipage}
\end{figure}

\begin{figure}[hbt!]
\begin{minipage}{1\linewidth}
\resizebox{2.4in}{!}{\includegraphics{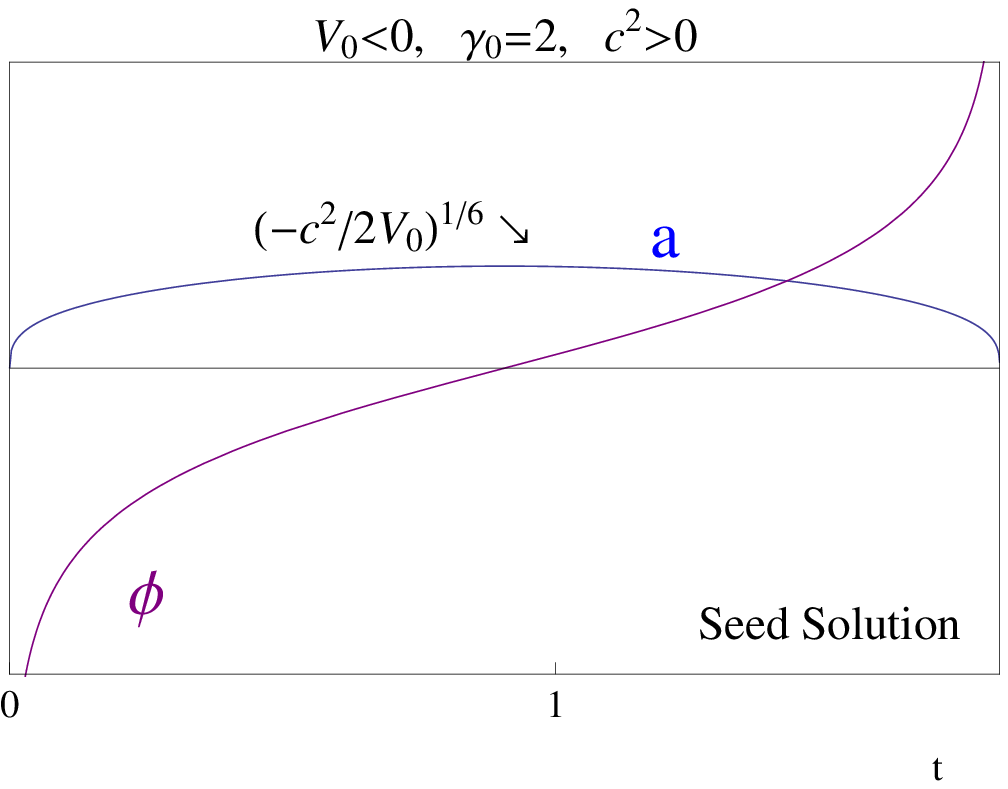}}\hskip0.05cm
\resizebox{2.4in}{!}{\includegraphics{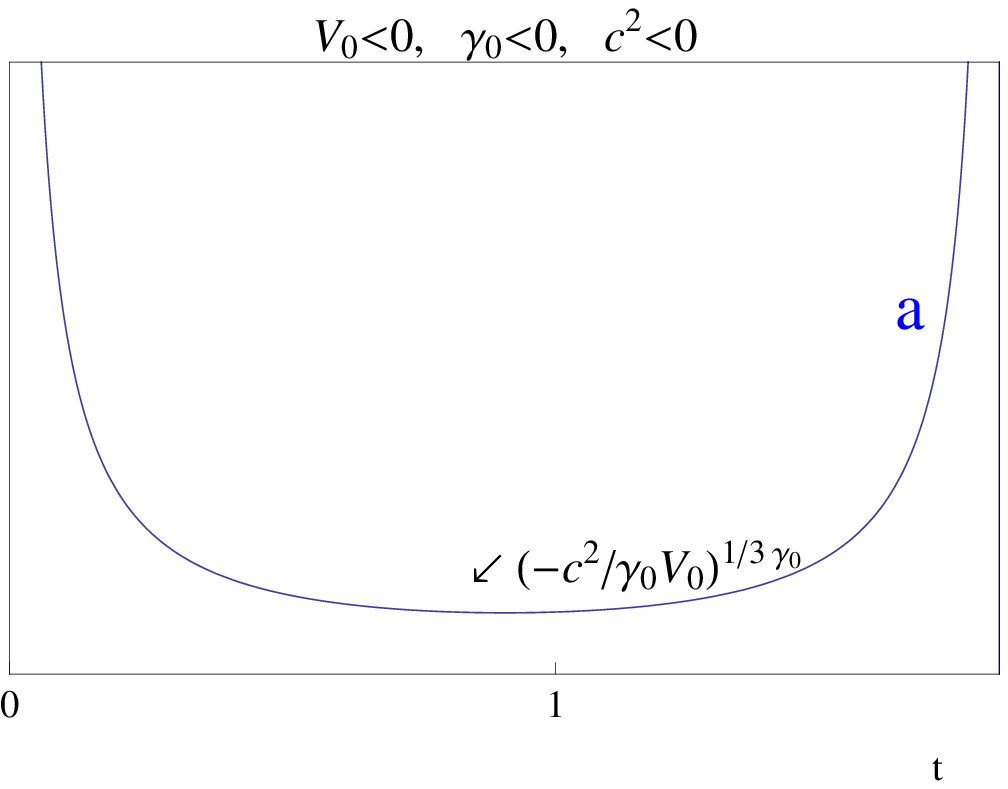}}\hskip0.05cm
\resizebox{2.4in}{!}{\includegraphics{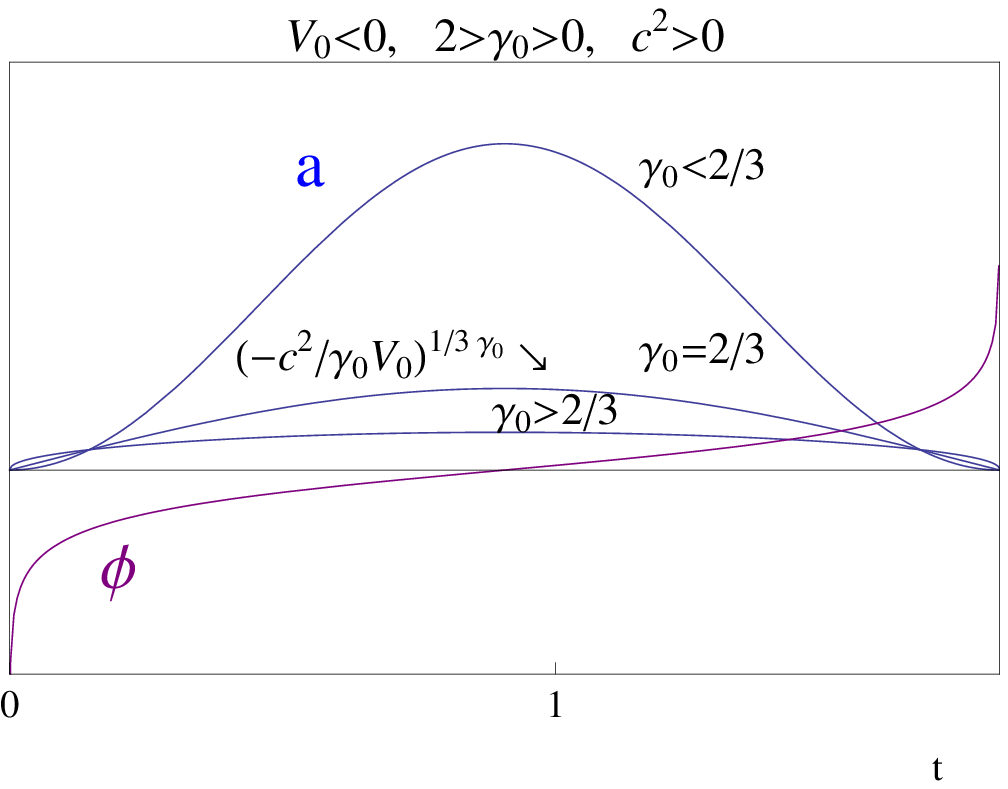}}\hskip0.05cm
\resizebox{2.4in}{!}{\includegraphics{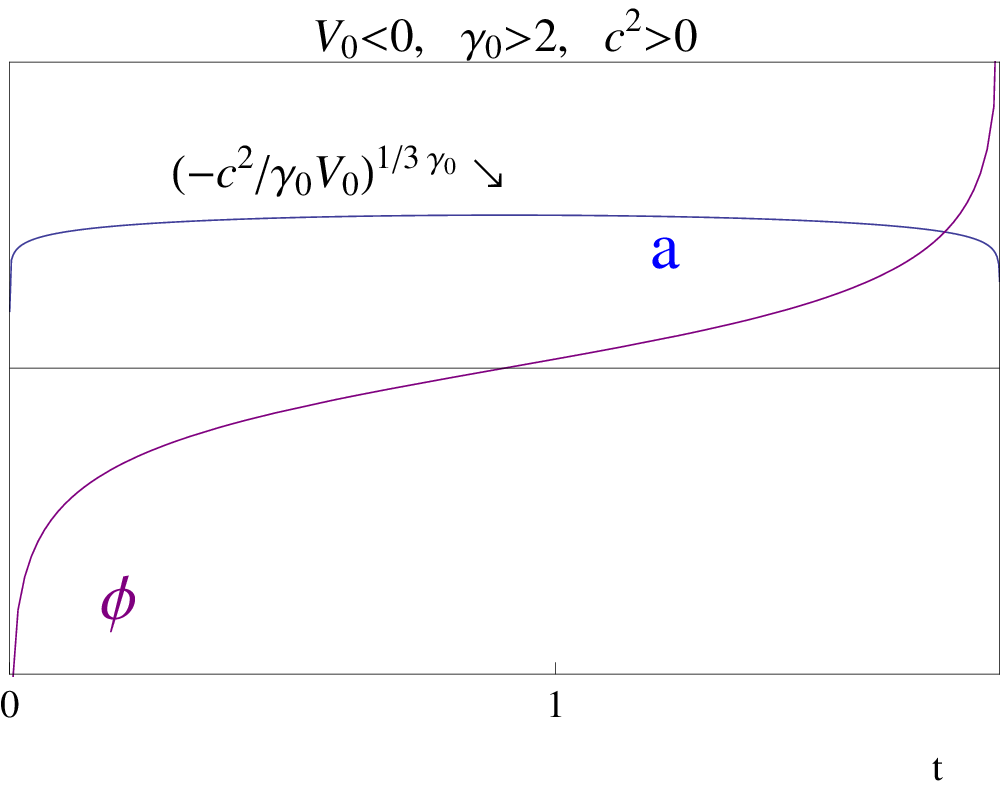}}\hskip0.05cm
\caption{\scriptsize{We plot $a(t)$ and $\phi(t)$ of the seed cosmology included in  ${\cal C}^{R}$ and the remaining cosmologies obtained by applying the FIT.}}
\label{F3}
\end{minipage}
\end{figure}
\subsubsection{Generating the set of solutions ${\cal B}$}

Applying the transformation rules (\ref{thn}), (\ref{tf})-(\ref{tv}) and (\ref{ns})-(\ref{tcs}) to the seed solution (\ref{ab})-(\ref{p.2b}), it turns into
\be
\label{abb}
\bar a=\left[\sqrt{\frac{-\bar{c}^2}{\bar\ga_0\bar{V_0}}}\,\,
\cosh{\frac{\sqrt{3\bar V_0\bar\ga_0^2}}{2}\,\,t}\right]^{2/3\bar\ga_0},
\ee
\be
\label{bp+-+}
\bar\phi=\frac{4}{\sqrt{-3\bar\ga_0}}\,\arctan{\exp{\frac{\sqrt{3\bar V_0\bar\ga_0^2}}{2}\,\,t}},
\ee
\be
\label{bpp+-+}
{\dot{\bar\phi}}^2=-\bar{V_0}\bar\ga_0\sin^2{\frac{\sqrt{3\bar\ga_0}}{2}\,\,i\bar\phi},
\ee
\be\n{vb}
\bar V=\bar{V_0}\left[\cos^2{\frac{\sqrt{3\bar\ga_0}}{2}\,\,i\bar\phi}+
\frac{\bar{\ga_0}}{2}\sin^2{\frac{\sqrt{3\bar\ga_0}}{2}\,\,i\bar\phi}\right].
\ee
Removing the bar in Eqs. (\ref{abb})-(\ref{vb}), we get the general solution of the system of equations (\ref{00s})-(\ref{pi}) included in the  set ${\cal B}$ or the ``solutions'' of the EKG equations (\ref{A})-(\ref{kg}) driven by the potential (\ref{vb}) which satisfy the condition $H=H(\phi)$.

The scale factor (\ref{abb}) describes a non singular universe having contracting and expanding phases. For $\ga_0>0$ and $c^2<0$, subset \bi, the scalar field is imaginary and the scale factor exhibits a bounce at $a_b=(-{c}^2/\ga_0{V_0})^{1/3\ga_0}$. Recently, it has been studied in detailed the conditions for  getting a bounce in FRW background within the context of  $f(R)$ models or  non-local actions with improved UV behavior, which  recover Einstein's general relativity in the IR; in particular, its connection with violation of unitarity  was analyzed in \cite{Mazumdar}. For $\ga_0<0$ and $c^2>0$,  subset \br, the scalar field is real and the scale factor reaches a maximum at $a_b$ [see Fig. (\ref{F2})]. It means that the FIT connect two different cosmological models; one with $V_0>0$, $\ga_0>0$, $\dot{\phi}^2<0$ and another with $V_0>0$, $\ga_0<0$, $\dot{\phi}^2>0$.

\subsubsection{Generating the set of solutions ${\cal C}$}

Applying the transformation rules (\ref{thn}), (\ref{tf})-(\ref{tv}), and (\ref{ns})-(\ref{tcs}) to the seed solution (\ref{ac})-(\ref{p.2c}), it turns into
\be
\bar a=\left[{\sqrt{\frac{-\,\bar c^2}{\bar{\ga_0}\bar V_{0}}}}\,\,\sin{\frac{\sqrt{-3\bar V_0\bar\ga_0^2}}{2}\,\,t}\right]^{2/3\bar\ga_0}, \label{ba-++}
\ee
\be
\n{bp-++}
\bar\phi=\frac{2}{\sqrt{3\bar\ga_0}}\,\ln{\tan{\frac{\sqrt{-3\bar V_0\bar\ga_0^2}}{4}\,\,t}},
\ee
\be
{\dot{\bar{\phi}}}^2=-\bar{V_{0}}\bar{\ga_0}\cosh^2{\frac{\sqrt{3\bar\ga_0}}{2}\,\,\bar{\phi}},  \label{bpp-++}
\ee
\be
\bar V=-\bar{V_0}\left[\sinh^2{\frac{\sqrt{3\bar\ga_0}}{2}\,\,\bar{\phi}}-\frac{\bar{\ga_0}}{2}\cosh^2{\frac{\sqrt{3\bar\ga_0}}{2}\,\,\bar{\phi}}\right].  \label{bVL-++}
\ee
Removing the bar in Eqs. (\ref{ba-++})-(\ref{bVL-++}), we get the general solution of the system of equations (\ref{00s})-(\ref{pi}) belonging to the set ${\cal C}$ or the ``solutions'' of the EKG equations (\ref{A})-(\ref{kg}) driven by the potential (\ref{bVL-++}) which satisfy the condition $H=H(\phi)$.

For $\ga_0>0$ and $c^2>0$, subset \cre, the scalar field is real and the scale factor (\ref{ba-++}) describes a universe with a finite time span that evolves from a Big Bang at $t=0$ and ends in a Big Crunch at $t=2\pi/\sqrt{-3\bar V_0\bar\ga_0^2}$ [see Fig. (\ref{F3})]. However, for  $\ga_0<0$ and $c^2<0$, subset \ci, the scalar field is imaginary and the universe ends in a Big Rip. Hence, the FIT connect two different cosmological models, one with $V_0<0$, $\ga_0>0$, $\dot{\phi}^2>0$ and another with $V_0<0$, $\ga_0<0$, $\dot{\phi}^2<0$.

\section{conclusion}

We have presented the FIS, that preserve the form of the EKG equations, for a spatially flat FRW universe and shown that the FIT have a Lie group structure. We have focused our investigations in the linear FIT generated by $\bar\ro=n^2\ro$ to show the duality between contracting cosmology and expanding one in a very intuitive fashion. We have found the exact solution for a perfect fluid with linear equation of state and for a scalar field driven by a potential that can be written as a linear function of the kinetic energy  density $V(\phi)=\frac{2-\ga_0}{2\ga_0}\,{\dot\phi}^{2}+V_0$.

We have explored the scalar field cosmology with two different seed solutions defined by the choices: $(i)-$ $\ga_0=2$ and  $V_0=0$ that leads to  $V=0$, and $(ii)-$ $\ga_0=2$ that gives $V=V_0$, see Table (\ref{I}). In the former case,  we have found that the FIT transform the seed scale factor $a_{_{\mp}}=[\mp\sqrt{3c^2/2}\,\,t]^{1/3}$ into the power law solution $a_{_{\mp}}=[\mp\sqrt{3c^2/2}\,\,t]^{1/3\ga}$ and the seed vanishing potential into the exponential potential $V=\frac{2(2-\ga_0)}{3\ga_0^2}\,\,e^{-\sqrt{3\ga_0}\,\phi}$. The latter includes three different seed solutions according to the signs of $V_0$ and $\ga_0 c^2$, which are included in the sets \ar, \bi and \cre, [see Table (\ref{I})]. The related seed scale factors obtained for $V=V_0$ are given below:

\begin{enumerate}
\item $a_{_{\mp}}=[\mp\sqrt{c^2/2V_0}\,\sinh{\sqrt{3V_0}\,\,t}]^{1/3}$ $\subset$ \,\, \ar,$~~~$
[see Table (\ref{I}) and Fig. (\ref{F1})],

\item$a=[\sqrt{{-\,c^2}/{2V_0}}\,\cosh{\sqrt{3V_0}\,\,t}]^{1/3}$  $\subset$ \,\,  \bi,$~~~$
[see Table (\ref{I}) and Fig. (\ref{F2})],

\item $a=[\sqrt{-\,c^2/2V_0}\,\sin{\sqrt{-3V_0}\,\,t}]^{1/3}$  $\subset$ \,\,  \cre,$~~~~$
[see Table (\ref{I}) and Fig. (\ref{F3})].
\end{enumerate}
For illustration purposes,  we have shown how  FIT generate new cosmologies from a seed one. Their scale factors and potentials are listed below:
\begin{enumerate}
\item $a_{_{\mp}}=[\mp\sqrt{c^2/\ga_0 V_0}\,\,\sinh{\sqrt{3 V_0\ga_0^2}\,\,t/2}]^{2/3\ga_0}$,\,
$V={V_0}[\cosh^2{\sqrt{3\ga_0}\,\,\phi/2}-{2^{-1}\ga_0}\sinh^2{\sqrt{3\ga_0}\,\,\phi/2}]$
are included in ${\cal A}$ [see Table (\ref{I}) and Fig. (\ref{F1})],

\item $a=[\sqrt{-c^2/\ga_0 V_0}\,\,\cosh{\sqrt{3V_0\ga_0^2}\,\,t/2}]^{2/3\ga_0}~$,\,
$V={V_0}[\cos^2{\sqrt{3\ga_0}\,\,i\phi/2}+2^{-1}\ga_0\sin^2{\sqrt{3\ga_0}\,\,i\phi/{2}}]$\,\,
are included in ${\cal B}$ [see Table (\ref{I}) and  Fig. (\ref{F2})],

\item  $a=[{\sqrt{-c^2/\ga_0 V_0}}\,\,\sin{\sqrt{-3V_0\ga_0^2}\,\,t/{2}}]^{2/3\ga_0}~$, \, $V=-V_0[\sinh^2{{\sqrt{3\ga_0}\,\,\phi/2}}-2^{-1}{\ga_0}\cosh^2{\sqrt{3\ga_0}\,\,{\phi}/{2}}]$ are included in ${\cal C}$ [see Table (\ref{I}) and Fig. (\ref{F3})].
\end{enumerate}

Summarizing, we have exploited  the Lie group structure of the FIS contained in the KGE equations and used a linear representation of this group, generated by the transformation $\bar\ro=n^2\ro$, to apply the induced FIT to a seed cosmology and have shown how to obtain a large set of  cosmologies driven by the scalar field, avoiding the direct integration computation of the EKG field equations.

\newpage

\section*{APPENDIX}

{ Let us now  assume  that we have a configuration of $n$ homogeneous scalar fields $\phi_{i}$ which do not interact between them, but are driven by  a sum of $n$ exponential potential $V_{i}=V_{0i}e^{-k_{i}\phi_{i}}$ so that the EKG equations read
\be
\label{AR}
3H^2=\sum_{i}{\frac{1}{2}\dot{\phi}^2_{i}+V_{0i}e^{-k_{i}\phi_{i}}},
\ee
\be
\n{kgR}
\ddot\phi_{i}+3H\dot\phi_{i}-k_iV_{0i}e^{-k_{i}{\phi_{i}}}=0.
\ee
From now on, we consider a simplified scheme where the configuration of the $n$ scalar fields is replaced by the configuration of one scalar field in which  the constants $k_{i}$ and $V_{0i}$ satisfy the conditions $k_{1}=k_{2}=...=k_{n}=k$  and  $V_{01}=V_{02}=...=V_{0n}=V_{0}$.  Under these assumptions the asymptotic evolution of all scalar fields tends to a common limit, meaning that the special scalar field configuration in which all scalar fields are equal is a late-time attractor. Hence, we may take  $\phi_{1}=\phi_{2}=...=\phi_{n}=\phi$ and $V_{1}=V_{2}=...=V_{n}\equiv V=V_{0}e^{-k\phi}$, then the EKG equation become
\be
\label{AR'}
3H^2=n \left[ \frac{1}{2}\dot{\phi}^2+V_{0}e^{-k\phi} \right],
\ee
\be
\n{kgR'}
\ddot\phi+3H\dot\phi-kV_{0}e^{-k\phi}=0.
\ee
They have a power-law solution $a_{n} \propto t^{2n/k^{2}}$, which inflates at all times when $k^{2}<2n$. Thus, the fields cooperate  to make inflation more likely.

Expanding cosmologies are clearly described by the Hubble parameter and the deceleration parameter $q (t)=-H^{-2}(\ddot{a}/a)$ that transforms as
\be
\n{tq}
\bar q+1=\left(\frac{\rho}{\bar\rho}\right)^{3/2}\frac{d\bar\rho}{d\rho}\,
(q+1),
\ee
\no under the FIT (\ref{tr})-(\ref{tp+r}). In the case we are considering the FIT is generated by the $\bar{\rho}(\rho)=n^{2}\rho$, in particular $\bar{\ro}$ and $\rho$ could be identified with the energy densities corresponding to  the fields $\bar{\phi}$ (associated with the $n$-field configuration) and $\phi$ (associated with the one-field configuration) respectively, and the deceleration parameter $q$ transforms as
\bn{tqs}
\bar q=-1+\frac{1}{n}(1+q),
\ee
while $\bar H=nH$ and $\bar a=a^n$. This shows that an expanding universe with a positive deceleration parameter transforms into an accelerated one $\bar q\to -1$, enhanced inflation \cite{LRM}, by taking $n$ large enough. In some sense, whenever $n>1$ it plays the role of increasing the energy density of the barred cosmology because $\bar{\rho}=n^{2}\rho$ and $\bar\ro>\ro$. Hence enhanced inflation, framed in the underlying structure of the FIS, generalizes the assisted inflation, where $n$ is the number of scalar fields, because here $n$ is a real number and  the enhanced inflation is achieved by the cooperative effects of adding energy density $n>1$ into the Friedmann equation (\ref{00s}) instead of adding scalar fields.}

\acknowledgments
We are grateful with the referee for useful comments that helped improve the article.
L.P.C thanks  the University of Buenos Aires under Project No. 20020100100147 and the Consejo Nacional de Investigaciones Cient\'{\i}ficas y T\' ecnicas (CONICET) under Project PIP 114-200801-00328. M.G.R and I.S.G are partially   supported by CONICET.


\end{document}